\documentclass[fleqn,usenatbib]{mnras}

\usepackage{newtxtext,newtxmath}

\usepackage[T1]{fontenc}

\DeclareRobustCommand{\VAN}[3]{#2}
\let\VANthebibliography\thebibliography
\def\thebibliography{\DeclareRobustCommand{\VAN}[3]{##3}\VANthebibliography}

\usepackage{graphicx}	
\usepackage{amsmath}	

\title[Study of correlated variability in Mkn 478]{Correlated variability of the reflection fraction with the X-ray flux and spectral index for Mkn 478}

\author[Barua et al.]{
Samuzal Barua,$^{1}$\thanks{E-mail: samuzal.barua@gmail.com}
V. Jithesh,$^{2}$
Ranjeev Misra,$^{3}$\thanks{E-mail: rmisra@iucaa.in}
Biman J Medhi,$^{1}$
Oluwashina Adegoke${^4}$
\\
$^{1}$Department of Physics, Gauhati University, Jalukbari, Guwahati 781014, Assam, India\\
$^{2}$Department of Physics and Electronics, CHRIST (Deemed to be University), Hosur Main Road, Bengaluru 560029, India\\
$^{3}$Inter-University Centre for Astronomy and Astrophysics (IUCAA), PB No. 4, Ganeshkhind, Pune 411007, India\\
$^{4}$Cahill Center for Astronomy and Astrophysics, California Institute of Technology, Pasadena, CA 91125, USAOluwashina Adegoke\\
}

\date{Accepted XXX. Received YYY; in original form ZZZ}

\pubyear{2021}

\begin{document}
\label{firstpage}
\pagerange{\pageref{firstpage}--\pageref{lastpage}}
\maketitle

\begin{abstract}
The X-ray spectrum of Mkn 478 is known to be dominated by  a strong soft excess which can be described using relativistic blurred reflection. Using observations from {\it XMM-Newton}, {\it AstroSat} and {\it Swift}, we show that for the long-term ($\sim$ years) and intermediate-term (days to months) variability, the reflection fraction is anti-correlated with the flux and spectral index, which implies that the variability is due to the  hard X-ray producing corona moving closer to and further from the black hole. Using flux-resolved spectroscopy of the {\it XMM-Newton} data, we show that the reflection fraction has the same behaviour with flux and index on short time-scales of hours. The results indicate that both the long-term and short-term variability of the source is determined by the same physical mechanism of strong gravitational light bending causing enhanced reflection and low flux as the corona moves closer to the black hole.         
\end{abstract}

\begin{keywords}
 black hole physics -- accretion discs -- galaxies: active -- galaxies: Seyfert - X-rays: individual: Mkn 478.
\end{keywords}



\section{Introduction}

Narrow-line Seyfert 1 (NLS1) galaxies are identified as a specific group of Active Galactic Nuclei (AGN) based on their  emission line properties, such as a narrow H$\beta$ line with typical FWHM < 2000  km s$^{-1}$, strong Fe II line and weak [O III] lines \citep{Osterbrock1985, Goodrich1989}. NLS1 have luminosities comparable to the Eddington limit and show strong variability in the soft X-ray regime \citep{Brandt1998}, which can provide information about the central black hole activity.

 The X-ray emission of AGN is believed to arise from the Compton up-scattering of   photons from a disc by electrons in a hot corona  \citep{Haardt1991, Pounds1995, Boller1996, Fabian2015, Adegoke2019}. A fraction of this  primary emission reflects back off the disc, giving reflection signatures. These reprocessed features typically consists of an iron fluorescence K shell emission peaking at around 6.5 keV and a Compton reflection hump at around 20--100 keV \citep{George1991, Turner1999, Ballantyne2001}. The iron line profile gets broadened due to relativistic effects in the vicinity of the black hole \citep{Fabian2000, Reynolds2003, Dauser2012} and hence can provide information about the black hole spin \citep{Laor1991}, physics of the accreting material under extreme relativistic effects \citep{Fabian2009}, and geometry of the primary X-ray source \citep{Wilkins2014}.

The strength of reflection can be quantified  by  measuring the reflection fraction which is usually defined as the ratio of reflected flux to the incident  flux. For a significant number of AGN, the reflection fraction has been observed to correlate with the photon index \citep{Malzac2001, Dadina2008, Molina2009, Reynolds2012, Balokovi2015, Boissay2016, Qiao2017, Zappacosta2018}. Recently, \citet{Ezhikode2020} have studied the correlation behaviour for 14 Seyferts and found the similar behaviour that the reflection fraction positively correlates with the spectral index. The observed correlation can be explained in terms of the accretion disc-corona geometry. If the corona subtends a larger solid angle with respect to the disc, then the number of soft photons entering the corona increases leading to a steepening of the photon index. The larger solid angle would also lead to higher reflection and hence the reflection fraction and index are correlated \citep{Zdziarski1999}. AGN also reveal softer-when-brighter behaviour when the photon index is seen to be positively correlated with the X-ray flux or the Eddington ratio. This has also been understood as the result of the Compton up-scattering process, where increase in the source flux leads to the softening of the X-ray spectrum along with an increase in flux \citep[e.g.][]{Zdziarski2003, Sobolewska2009, Boissay2016, Huang2020}. It should be noted that these results are for the cases when the reflection fraction is of the order of or less than unity.

Individual observations of  sources have revealed high reflection fraction (exceeding unity) in the low flux states of AGN (Seyferts) MCG-6-30-15 \citep{Fabian2003}, 1H 0707-495 \citep{Fabian2004}, 1H0419-577  \citep{Fabian2005} and a black hole binary XTE J1650-500 \citep{Rossi2005, Reis2013}. This has been interpreted as the X-ray source being very close to the black hole and hence its emission is gravitationally red-shifted resulting in the low flux, but strong light bending effects enhance the reflection fraction leading to a reflection dominated spectrum \citep{Fabian2006, Miniutti2003, Miniutti2004}. In this scenario, as the X-ray source moves closer to the black hole, the primary flux should decrease while the reflection fraction should increase, leading to an anti-correlation between the two.
Using multiple {\it XMM-Newton} observations of the  NLS1 1H 0419-577,
\citet{Fabian2005} showed that the flux and the reflection fraction are anti-correlated for this reflection dominated source (c.f. Fig. 7 of \citet{Fabian2005}).

The multi-epoch {\it XMM-Newton} observations have revealed that the X-ray spectra of the NLS1 Mkn 478 is dominated by the  blurred reflection component \citep{Zoghbi2008}. \citet{Waddell2019} conducted a detailed analysis of the {\it XMM-Newton} and {\it Suzaku} data-sets and showed that the X-ray spectra can be adequately described to be dominated by blurred reflection, which  provided an estimate of the black hole spin to be $\sim$0.98. Their study showed a similarity between the nature of this source and NLS1 1H 0419-577.

Here, we reanalyze the {\it XMM-Newton} observations of Mkn 478 and include new analysis of  {\it AstroSat} and multi-epoch {\it Swift} observations, to study the correlation between reflection fraction and spectral properties of this reflection dominated source over long (years) time-scale. Additionally, we perform flux-resolved spectroscopy of the {\it XMM-Newton} observations to see how the short-term (hours) correlations compare with long-term ones. 

The paper is organised as follows. In Section 2, we describe the observations and data reduction procedures. In Section 3, we present findings from  the spectral analyses of time-averaged and  flux-resolved spectra. We then discuss and interpret the results in Section 4.

\section{Observation and data reduction}

\begin{table*}
	\centering
	\caption{Observations from three  observatories used for spectroscopic analysis. In the third column the observations are named sequentially.}
	\label{tab:table1}
	\setlength{\tabcolsep}{7.0pt}
	\begin{tabular}{lcccccc} 
		\hline
		Satellite & Observation ID & Name & Observing time & Duration &  Count rate  & Energy range\\  
		
		          &                &       & (yyyy-mm-dd)    &     (ks) &  (c/s) &  (keV)\\
		
		\hline
		   {\it XMM-Newton} & 0107660201 & Data1 & 2001-12-23 & 32.616 & 5.2 & 0.3--10 \\
		 & 0005010101 & Data2 & 2003-01-01 &27.751 & 5.5 & \\
		 & 0005010301 & Data3 &2003-01-07 &26.453 & 2.86  &\\
		 & 0801510101 & Data4 &2017-06-30 &135.300& 3.15 & \\
		{\it AstroSat}/SXT &A05\_175T01\_9000002894  & Data5 & 2019-05-07 & 30 & 0.116 & 0.3--6 \\
		{\it AstroSat}/LAXPC & --      &       &   & 54 & 0.31  & 4--10\\
		{\it Swift} & 00035903003 & Data6 &2007-09-26 & 1.523 & 2.32 & 0.2--4 \\
		      & 00091339001 & Data7 &2013-01-13& 0.539 & 2.738 &  \\
		      & 00091339002 & Data8&2013-02-28 & 0.499 & 1.61 & \\
		      & 00091339003 & Data9&2013-03-02 & 0.842 & 3.11 & \\
		      & 00093091001 & Data10&2017-09-20 & 0.890 & 2.56 & \\  
		      & 00095086001 & Data11 & 2019-09-16 &0.9076 & 0.29 & \\
		      & 00095896001 & Data12&2021-10-03 & 0.943 & 3.33 & \\	
		      	\hline
	\end{tabular}
\end{table*}

\subsection{\it XMM-Newton}
Mkn 478 has been observed on five occasions with {\it XMM-Newton} \citep{Jansen2001} satellite since 2001. One of the observations (Obs ID : 0005010201) was of poor quality \citep{Waddell2019} and hence was not considered for the analysis. The details for the four observations are given in Table~\ref{tab:table1}. The {\it XMM-Newton} data files from EPIC-pn camera were processed by following standard data reduction procedure with the Science Analysis System ({\sc sas} v.17.0.0). We used the {\sc evselect} task to generate the event files, then cleaned the files for high background  flaring. From that, the good time intervals (gti) files were created using the {\sc evselect}. From the cleaned event files, we extracted the source and background events from  circular regions of radii 35 and 40 arcseconds, respectively. These regions were used for generating the source light curve, background light curve, source spectrum and background spectrum. To obtain a background-subtracted lightcurve in the 0.3--10 keV band, we used the {\sc epiclccorr} task. We employed {\sc arfgen} and {\sc rmfgen} tasks for generating arf and rmf files. Finally, we grouped the spectrum using {\sc grppha} and used it for the spectral analysis.

\begin{figure*}
	\includegraphics[width=8.85cm, height=7cm]{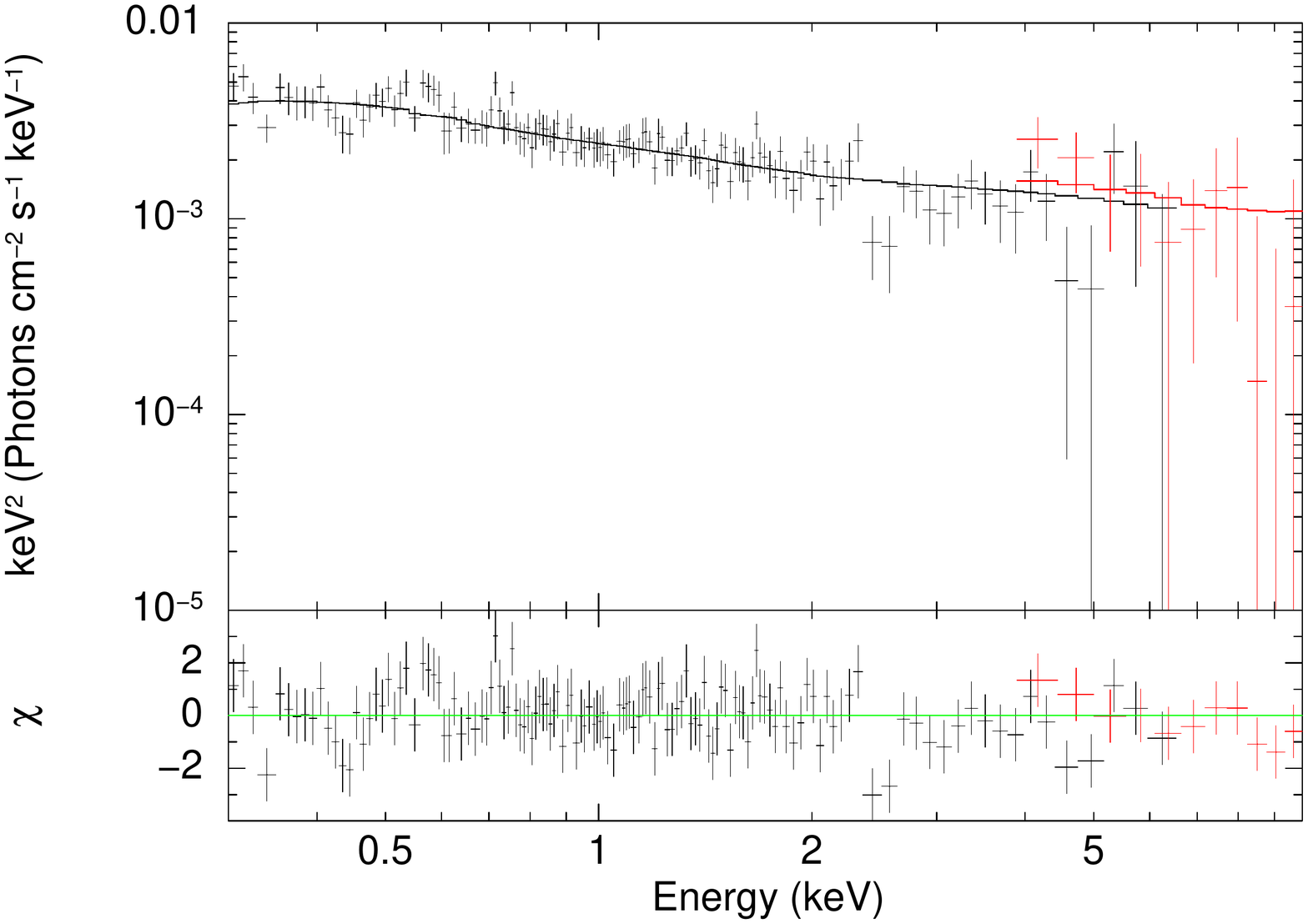}
	\hspace{0cm}
	\includegraphics[width=8.85cm, height=7cm]{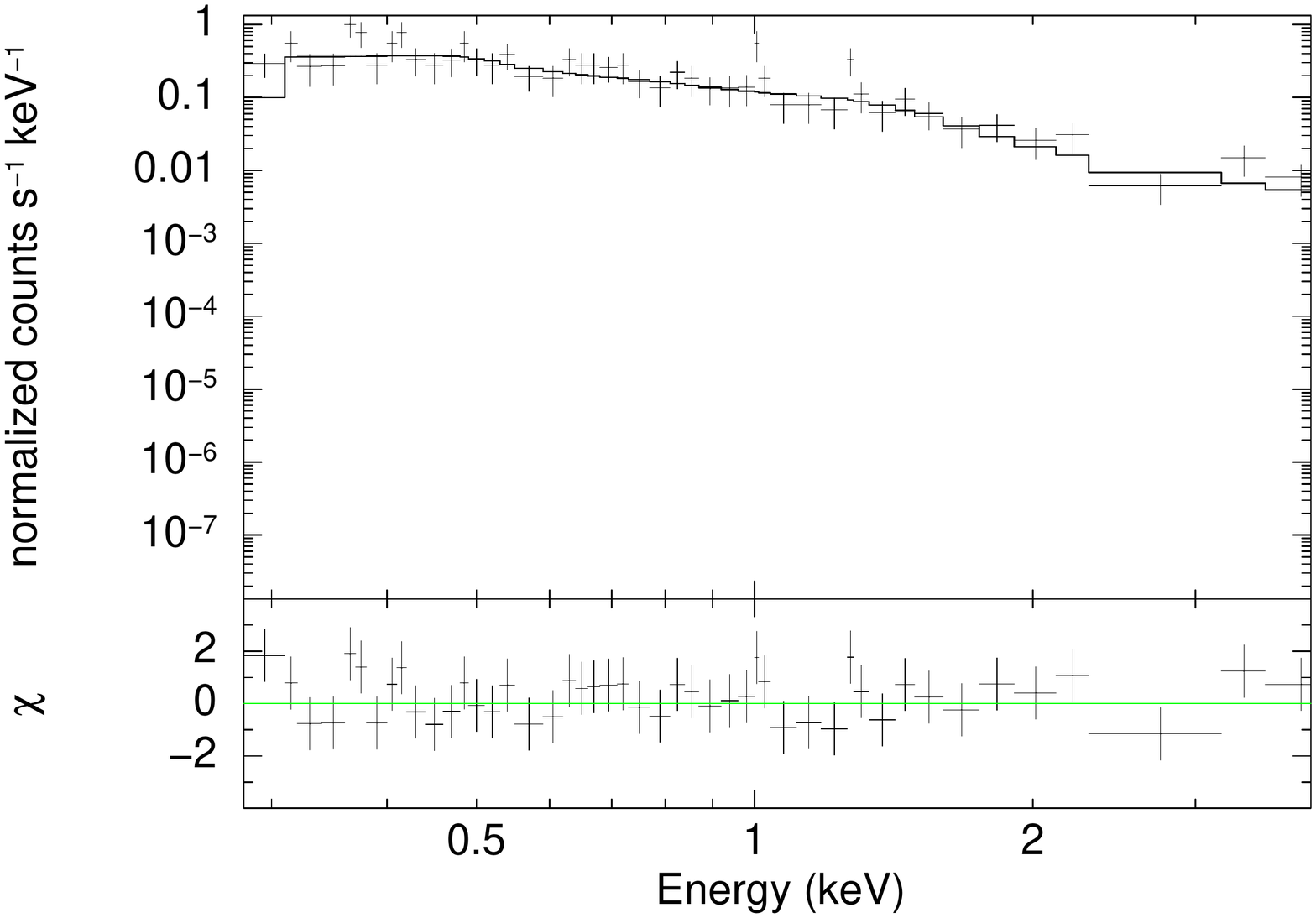}
	\vspace{-0.8cm}
	\caption{ Left panel represents the {\it AstroSat} spectra from the simultaneous fits of 0.3--6 keV SXT and 4--10 keV LAXPC observations and the right panel shows 0.2-4 keV spectra from the {\it Swift} XRT observation (Obs ID : 00095086001). Both {\it AstroSat} and {\it Swift} spectra are fitted with relativistic reflection model with parameters listed in Tables \ref{tab:table3} and \ref{tab:table4}. }
	\label{fig:figure1}
\end{figure*}

\subsection{\it AstroSat}

{\it AstroSat} \citep{Singh2014} observed Mkn 478 with the Soft X-ray Telescope (SXT) and the Large Area X-ray Proportional Counter (LAXPC) instruments simultaneously on 2019 May 7. The SXT and LAXPC observations were conducted for 30 and 54 ks exposure time, respectively. SXT is an imaging telescope that operates in the 0.3--8 keV energy range. It has a spectral resolution of $\sim$ 150 eV at 6 keV and the on-axis full width at half maximum (FWHM) of the point spread function (PSF) in the focal plane is $2'$. For this observation, we used the Photon Counting (PC) mode of SXT. We used {\sc sxtpipeline} software (version 1.4b) to process the Level-1 data and produced the cleaned event file list for each orbit. The individual event files were merged using {\sc sxtevtmerger} tool. We used the merged event file to extract the final data products using {\sc xselect}.

LAXPC is a non-imaging telescope operating in the 3--80 keV energy band. It consists of three proportional counters, LAXPC10, LAXPC20 and LAXPC30 with a  temporal resolution of 10 micro-second. During the observation, LAXPC30 was not in working condition and LAXPC10 was not suitable for spectroscopic analysis because of its low gain and response. Thus, we used LAXPC20 data for our spectral analysis. The LEVEL-1 LAXPC20 data was processed with the software {\sc LaxpcSoft}, and the standard tools {\it laxpc\_make\_event} and {\it laxpc\_make\_gti} were used to extract the event data and GTI files, respectively. The level 2 event file was used to extract the spectrum.

For further analysis the spectra from both {\it AstroSat} and {\it Swift} were grouped using grppha tool so that each spectral bin contains at least 20 counts. The energy band chosen from {\it AstroSat} is 0.3--6 keV for SXT and 4--10 keV for LAXPC, whereas 0.2--4 keV is for {\it Swift} XRT. The count rate beyond these energy ranges were found to be negligible and hence were excluded from the analysis.

\subsection{\it Swift}
Mkn 478 has also been observed by the {\it Neil Gehrels Swift Observatory} \cite[{\it Swift};][]{Gehrels2004} on 18 occasions during the 2006--2021 period. Of these, we used seven {\it Swift}/XRT observations in this work as presented in Table~\ref{tab:table1}.  The other observations were found to not be suitable for spectroscopic analysis since the counts were low and hence were excluded from the analysis. We employed the standard data reduction technique for processing and screening {\it Swift}/XRT data files using {\sc xrtpipeline} (version: 0.13.5). For all these observations PC mode data were taken into consideration. The source and background events were extracted from a circular region of radius 47 arcseconds and we used the standard grade filtering 0--12. We created the response matrix file using the {\sc xrtmkarf} tool.

\subsection{\it Suzaku}
Mkn 478 was also observed for $\sim$ 85 ks by {\it Suzaku} satellite with XIS camera. But the limited X-ray energy range (0.7--8 keV; see \citet{Waddell2019}) would not allow to constrain the reflection parameters and hence the {\it Suzaku} observation was not considered in this work.

\section{Spectral fitting}

\subsection{Time-averaged spectra}

The {\it XMM-Newton} data has been analyzed in detail by  \citet{Waddell2019} and here we follow their prescription. They used two reflection models {\tt relxill}\footnote{\url{http://www.sternwarte.uni-erlangen.de/~dauser/research/relxill/}} \citep{Dauser2014, Garcia2014} as well as {\tt reflionx} \citep{Ross2005} to characterise the {\it XMM-Newton} spectra. They preferred the fit provided by reflionx since it required a lower reflection fraction. However, in this work we use the more up to date {\tt relxill} model which includes the latest radiative process treatment and includes the possibility that the reflecting disc is highly ionized. The disk ionisation is parameterised by the ionisation parameter given by $\xi = 4 \pi F/n$, where $F$ is the irradiating flux and $n$ is the disk density \citep{Garcia2013}. For a reflection spectrum the {\tt relxill} allows to estimate $\xi$ from 0 (neutral) to 4.7 (heavily ionised) in a logarithmic scale. Since \citet{Waddell2019} reported that there is no significant detection of an additional powerlaw component, we did not include this component in the spectral fitting. We did not find any significant evidence for any narrow line feature at around 6.7 keV with $\Delta \chi^2$ decreasing by a maximum of 1.5 for Data3 and less for the other spectra. Moreover, the inclusion of the narrow line had little impact on the best fit values of the other spectral parameters. Hence we have not included the feature. 
Some parameters were fixed to typical values, while some others were tied between observations, and the rest were allowed to vary independently. The inner ($R_{in}$) and outer radii ($R_{out}$) were fixed to the ISCO value and 400$r_g$ respectively. Absorption was taken into account using the {\tt tbabs} \citep{Wilms2000} with the column density fixed to  $\rm 1.08 \times 10^{20} cm^{-2}$ \citep{Willingale2013}.

While {\tt relxill} allows for a broken power-law model for the emissivity index, here for simplicity we used only a single power-law i.e. the index $q = q_{in}= q_{out}$. The model requires as parameters, the normalized value of black hole spin ($a$), the inclination angle of the disc to the observer ($\theta$), and iron abundance ($A_{Fe}$). Since these should not vary for the system, they were tied for the different spectra. For the spectral fitting we used background subtracted spectra for all observations in XSPEC (version: 12.11.1) \citep{Arnaud1996}. The unabsorbed  0.3--10 keV energy flux was estimated using the XSPEC model {\tt cflux}. Parameter errors were estimated by undertaking the   MCMC analysis using the {\sc XSPEC\_EMCEE}\footnote{\url{https://github.com/jeremysanders/xspec_emcee}} code build-up by Jeremy Sanders. It is a pure Python based Goodman $\&$ Weare's Affine Invariant Markov chain Monte Carlo (MCMC) Ensemble sampler. Here, we used 50 walkers with 15000 iterations and burned the first 1000. The quoted errors are estimated at the 1$\sigma$ confidence interval. As expected, the spectral fitting results are approximately similar to those obtained by \citet{Waddell2019} and Table~\ref{tab:table2} lists the best-fit parameters with errors and the $\chi^2$ value.

\begin{table*}
	\centering
	\caption{ Best-fit parameters from the {\it XMM-Newton} time-averaged spectra. For all datasets outer disc radius $R_{out}$ and inner disc radius $R_{in}$ are frozen to 400$r_g$ and 1$r_g$, respectively whereas the break radius $R_{br}$ has been frozen to 6$r_g$. Spin $a$, inclination angle $\theta$ and iron abundance $A_{fe}$ are tied between observations. The superscript 1 implies that the values are tied to Data1. The flux is unabsorbed flux in 0.3--10 keV.}
	\label{tab:table2}
	\begin{tabular}{lccccccccc} 
		\hline
		Observation & $\Gamma$ & $q$ & $a$ & $\theta$ & log$\xi$ & $A_{fe}$ & R & log$F$ & $\chi^2$/d.o.f\\

		&        &    &     & (degree) & log(erg cm $\rm s^{-1}$) & Fe/solar &    & log(erg $\rm cm^{-2}$ $\rm s^{-1}$) & \\
		
		\hline
		Data1 & $2.51_{-0.004}^{+0.01}$ & $6.74_{-0.27}^{+0.22}$ & $0.98_{-0.006}^{+0.003}$ & $33.18_{-3.48}^{+2.01}$ & $3.02_{-0.04}^{+0.02}$ & $0.93_{-0.027}^{+0.097}$ & $3.15_{-0.96}^{+0.21}$& $-10.88_{-0.02}^{+0.02}$ &  1136/1085 \\\\
		
		Data2 & $2.49_{-0.01}^{+0.02}$ & $6.92_{-0.32}^{+0.16}$ & $\rm 0.98^1$ & $\rm 33.18^1$ & $2.80_{-0.16}^{+0.03}$ & $\rm 0.93^1$ & $2.40_{-0.19}^{+0.09}$  & $-10.85_{-0.02}^{+0.02}$& -- \\\\
		
		
		Data3 & $2.45_{-0.02}^{+0.03}$ & $7.17_{-0.41}^{+0.27}$ & $0.98^1$ & $33.18^1$ & $2.80_{-0.16}^{+0.03}$ & $0.93^1$ & $2.74_{-0.45}^{+0.22}$ & $-11.12_{-0.03}^{+0.03}$ & --  \\\\
		
		Data4 & $2.43_{-0.01}^{+0.01}$ & $7.78_{-0.036}^{+0.21}$ & $0.98^1$ & $33.18^1$ & $2.90_{-0.06}^{+0.03}$ & $0.93^1$ & $3.15_{-0.35}^{+0.07}$ & $-11.09_{-0.01}^{+0.01}$ & --\\\\

		\hline
	\end{tabular}
\end{table*}


\begin{table*}
	\centering
	\caption{ Best-fit parameters from the {\it AstroSat} SXT and LAXPC spectra. All parameters from LAXPC are tied to those from SXT. Frozen parameters  values are same as those obtained for {\it XMM-Newton} fits and are represented by superscript f. During the analysis the multiplicative constant is fixed at 1 for SXT, and left free for LAXPC. The flux is unabsorbed flux in 0.3--10 keV band.}

	\label{tab:table3}
	\setlength{\tabcolsep}{7pt}
	\begin{tabular}{lcccccccccc} 
		\hline
		Observation & Scale factor & $\Gamma$ & $q$ & $a$ & $\theta$ & log$\xi$ & $A_{fe}$ & R & log$F$ & $\chi^2$/d.o.f\\

		&         &      &    &     & (degree) & log(erg cm $\rm s^{-1}$) & Fe/solar &    & log(erg $\rm cm^{-2}$ $\rm s^{-1}$) & \\
		
		\hline
		
		Data5 & $1.47_{-0.35}^{+0.33}$ &  $2.45_{-0.06}^{+0.06}$ & < 3 & $\rm 0.98^f$ &$ \rm 33.18^f$ & $3.83_{-0.90}^{+0.04}$ & $\rm 0.93^f$ & $3.32_{-0.49}^{+0.51}$ & $-11.14_{-0.015}^{+0.011}$ & 130/137 \\\\
		

		\hline
	\end{tabular}
\end{table*}

\begin{table}
	\centering
	\caption{ Best-fit parameters from the {\it Swift} spectra. Some parameters, respectively, emissivity $q$, spin $a$, inclination $\theta$ and iron abundance $A_{fe}$ are fixed to those obtained from {\it XMM-Newton} fit. The outer radius $R_{out}$, inner radius $R_{in}$ and break radius $R_{br}$ are also fixed to the values used for the {\it XMM-Newton} fit. The flux is the unabsorbed flux in 0.3-10 keV band. }
	\label{tab:table4}
	\setlength{\tabcolsep}{3pt}
	\begin{tabular}{lcccc} 
		\hline
		Observation & $\Gamma$ & log$\xi$ & R & log$F$ \\

		&         &  log(erg cm $\rm s^{-1}$) &    & log(erg $\rm cm^{-2}$ $\rm s^{-1}$) \\
		\hline
		
		Data6 & $2.40_{-0.08}^{+0.02}$ & $2.04_{-0.17}^{+0.04}$ & $2.90_{-0.25}^{+0.24}$ & $-11.15_{-0.02}^{+0.06}$ \\\\
		
		Data7 & $2.50_{-0.036}^{+0.15}$ & $2.04^6$ & $2.1_{-0.23}^{+0.21}$ & $-11.07_{-0.06}^{+0.02}$ \\\\
		
		Data8& $2.60_{-0.03}^{+0.10}$& $2.04^6$ & $0.71_{-0.10}^{+0.09}$ & $-11.70_{-0.05}^{+0.07}$ \\\\
		
		Data9& $2.54_{-0.08}^{+0.09}$& $2.04^6$ & $>0.1$ &  $-10.93_{-0.32}^{+1.47}$ \\\\
		
		Data10& $2.56_{-0.017}^{+0.05}$& $2.04^6$ & $0.9_{-0.1}^{+0.09}$ & $-10.73_{-0.08}^{+0.03}$ \\\\
		
		Data11& $2.50_{-0.12}^{+0.12}$& $2.04^6$ & $1.82_{-0.18}^{+0.16}$ & $-10.95_{-0.1}^{+0.03}$ \\\\
		
		Data12& $2.55_{-0.06}^{+0.05}$& $2.04^6$ & $1.1_{-0.33}^{+0.33}$ & $-10.80_{-0.01}^{+0.03}$ \\
		\hline
		
		$\chi^2$/d.o.f& & & & 132/125\\

		\hline
	\end{tabular}
\end{table}

\begin{figure}
	\includegraphics[width=8.5cm, height=6cm]{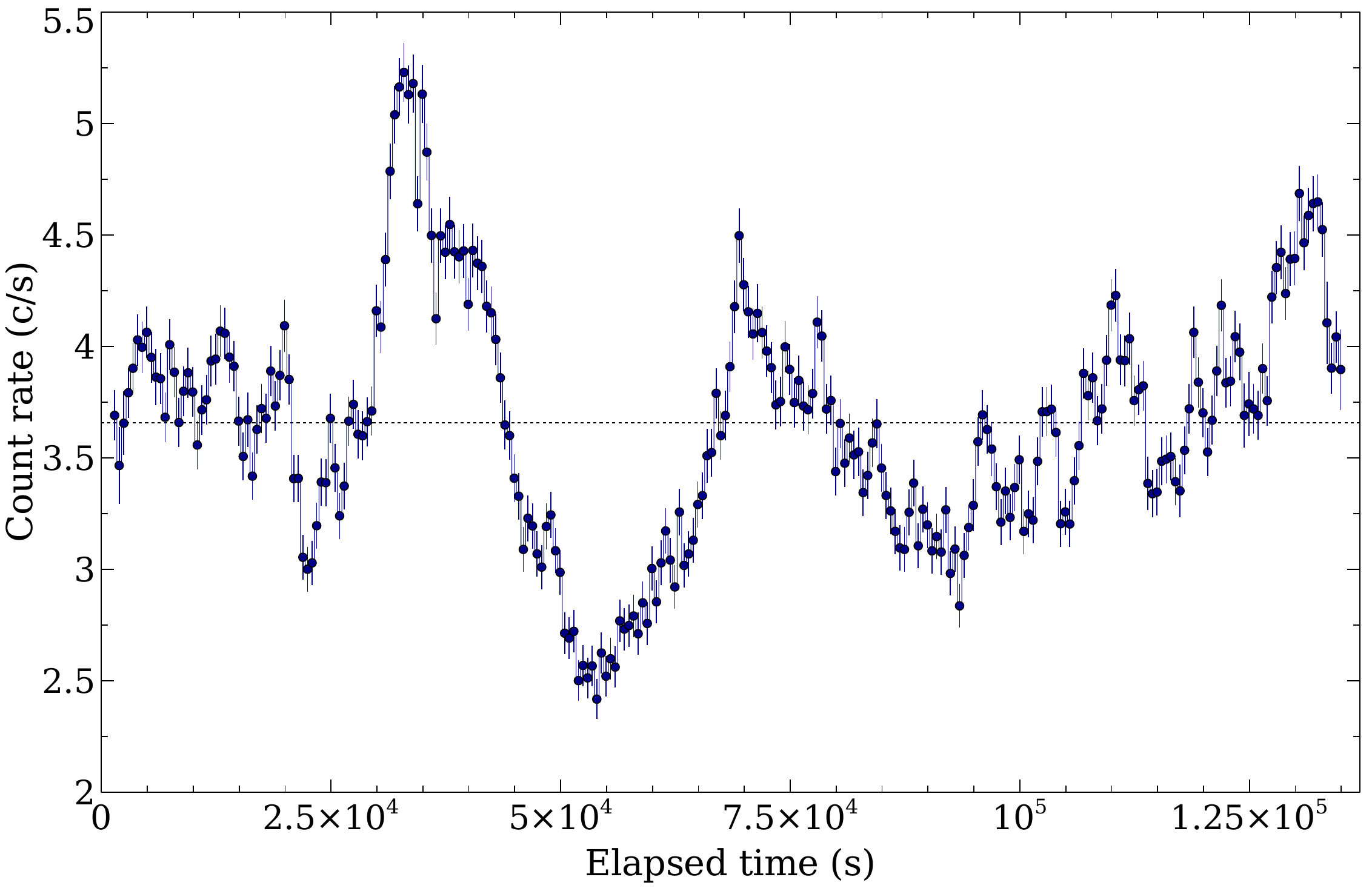}
	\caption{ Lightcurve extracted from Data4 with time bin size of 500 s. The horizontal line split the lightcurve into two states, where the threshold value is the average 0.2--10 keV count rate of 3.63 c/s. }
	\label{fig:figure2}
\end{figure}

We used the same model to fit  the {\it AstroSat} and {\it Swift} spectra. For {\it AstroSat} analysis, we fitted both SXT and LAXPC spectra simultaneously and used a constant  to take into account  the cross-calibration uncertainty between the two instruments.  The constant parameter is fixed to 1 for SXT and was allowed to vary for LAXPC which was constrained to $1.47_{-0.35}^{+0.33}$. The {\it AstroSat} spectrum yields a rather non-physical emissivity index of $< 3$, but this could be due to the poorer energy resolution and hence is not trustworthy. The black hole spin $a$, iron abundance $A_{fe}$ and inclination angle $\theta$ have been fixed to the values  obtained from {\it XMM-Newton} analysis and  the best-fit spectral parameters with error are given in Table~\ref{tab:table3}. The left panel of Figure \ref{fig:figure1} shows the AstroSat spectra and the best-fit model. The limited energy range and lower statistics of the {\it Swift} spectra did not allow for the emissivity index to be constrained and hence apart from the black hole spin, inclination angle and iron abundance, we had to also fix the index $q = 7$ as obtained from the  {\it XMM-Newton} analysis. The spectral parameters for the {\it Swift} analysis are listed in Table~\ref{tab:table4} and the right panel of  Figure \ref{fig:figure1} shows a representative spectrum along with the best-fit model.

\begin{table*}
	\centering
	\caption{Best-fit parameters from the {\it XMM-Newton} flux-resolved spectra. The low and high flux spectra from each observation are fitted simultaneously by adding a constant parameter, where the scale factor factor has been fixed at 1 for both the spectra. The analysis is similar to that for time-averaged spectra. Few parameters, such as spin $a$, iron abundance $A_{fe}$ and inclination $\theta$ have been fixed to those obtained from time-averaged fits. Other some parameters, inner radius $R_{in}$, break radius $R_{br}$, outer radius $R_{out}$ and emissivity index $q$ have been remained same to the time-averaged one. The flux is the unabsorbed flux in 0.3-10 keV band.}
	\label{tab:table5}
	\setlength{\tabcolsep}{6.5pt}
	\begin{tabular}{lcccccccccccc} 
		\hline
		Observation & Flux & $\Gamma$ & $q_{in}$ & $a$ & $\theta$ & log$\xi$ & $A_{fe}$ & R & log$F$ & $\chi^2$/d.o.f\\

		&         &      &    &     & (degree) & log(erg cm $\rm s^{-1}$) & Fe/solar &    & log(erg $\rm cm^{-2}$ $\rm s^{-1}$) & \\
		
		\hline
		Data1 & low & $2.496_{-0.016}^{+0.03}$ & $7.02_{-0.277}^{+0.377}$ & $\rm 0.98^f$ &$\rm 33.18^f$ & $3.01_{-0.03}^{+0.04}$ & $\rm 0.93^f$ & $2.458_{-0.35}^{+0.38}$& $-10.936_{-0.004}^{+0.004}$ &  535/531 \\\\
		
		& high & $2.54_{-0.01}^{+0.02}$ & $6.88_{-0.43}^{+0.34}$ & -- & -- & $2.95_{-0.06}^{+0.04}$ & -- & $1.5_{-0.15}^{+0.06}$  & $-10.839_{-0.004}^{+0.004}$& -- \\\\
		
		Data2 & low & $2.438_{-0.02}^{+0.03}$ & $7.33_{-0.40}^{+0.48}$ & $\rm 0.98^f$ & $\rm 33.18^f$ & $2.85_{-0.035}^{+0.065}$ & $\rm 0.93^f$ & $2.40_{-0.20}^{+0.27}$ & $-11.00_{-0.003}^{+0.003}$& 249/219\\\\
		
		& high & $2.50_{-0.03}^{+0.03}$ & $7.03_{-0.27}^{+0.31}$ & --& -- & $2.69_{-0.015}^{+0.08}$ & -- & $1.77_{-0.19}^{+0.31}$ & $-10.933_{-0.003}^{+0.003}$ & --  \\\\
		
		Data3 & low & $2.373_{-0.02}^{+0.025}$ & $7.25_{-0.62}^{+0.36}$ & $\rm 0.98^f$ & $\rm 33.18^f$ & $2.84_{-0.29}^{+0.03}$ & $\rm 0.93^f$ & $3.673_{-0.70}^{+0.37}$ & $-11.281_{-0.004}^{+0.004}$ & 517/510\\\\

		& high   & $2.47_{-0.025}^{+0.022}$ &$6.98_{-0.38}^{+0.29}$& -- &-- & $2.85_{-0.06}^{+0.18}$ &-- & $2.943_{-0.37}^{+0.36}$ & $-11.1965_{-0.004}^{+0.004}$ & -- \\\\
		
		Data4 & low & $2.421_{-0.02}^{+0.02}$ & $8.02_{-0.2}^{+0.40}$ & $\rm 0.98^f $ & $33.18^1$ & $3.0_{-0.03}^{+0.02}$ & $\rm 0.93^f$ & $3.966_{-1.13}^{+0.66}$ & $-11.27_{-0.002}^{+0.002}$& 327/293\\\\
		
		& high& $2.424_{-0.02}^{+0.013}$ & $7.24_{-0.17}^{+0.20}$ &  -- & -- &  $3.0_{-0.02}^{+0.014}$ & -- & $3.469_{-0.65}^{+0.43}$ & $-11.164_{-0.002}^{+0.002}$ & --\\
		
		\hline
	\end{tabular}
\end{table*}

\begin{figure*}
	\begin{minipage}{.48\textwidth}
		\includegraphics[width=7.5cm, height=6.2cm]{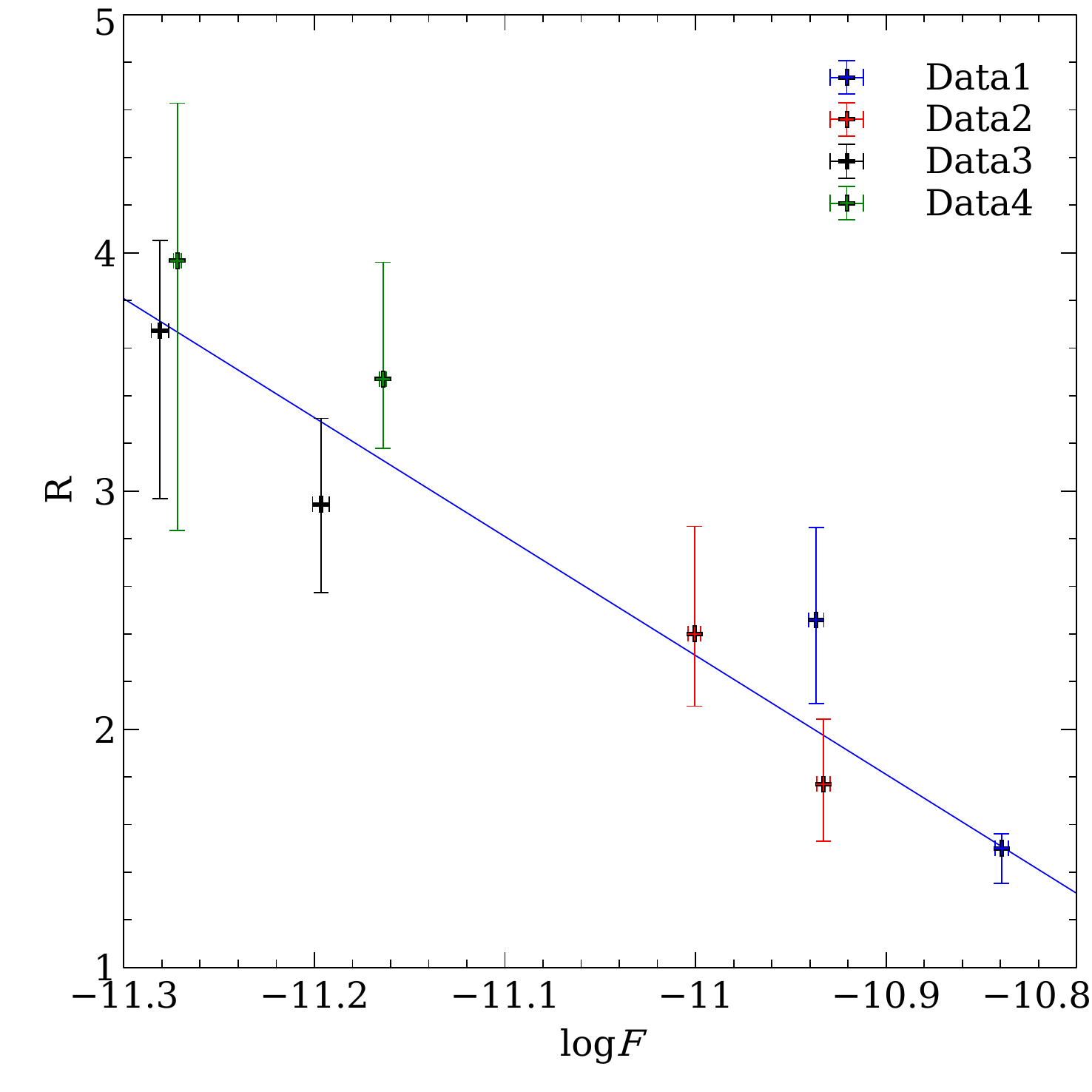}
		\caption{ Plot for reflection fraction (R) versus 0.3--10 keV unabsorbed flux from low and high flux spectra of different {\it XMM-Newton} observations. Different colors represent different observations with low and high flux spectra from each. The blue line is the best fit line obtained from the function $a + bx$, where $a = -52.5 \pm 6.51$, $b = -5 \pm 0.6$ and reduced $\chi^2 = 0.7$.}
		\label{fig:figure3}
	\end{minipage}
	\hspace{0.4cm}
	\begin{minipage}{.44\textwidth}
		\includegraphics[width=7.5cm, height=6.2cm]{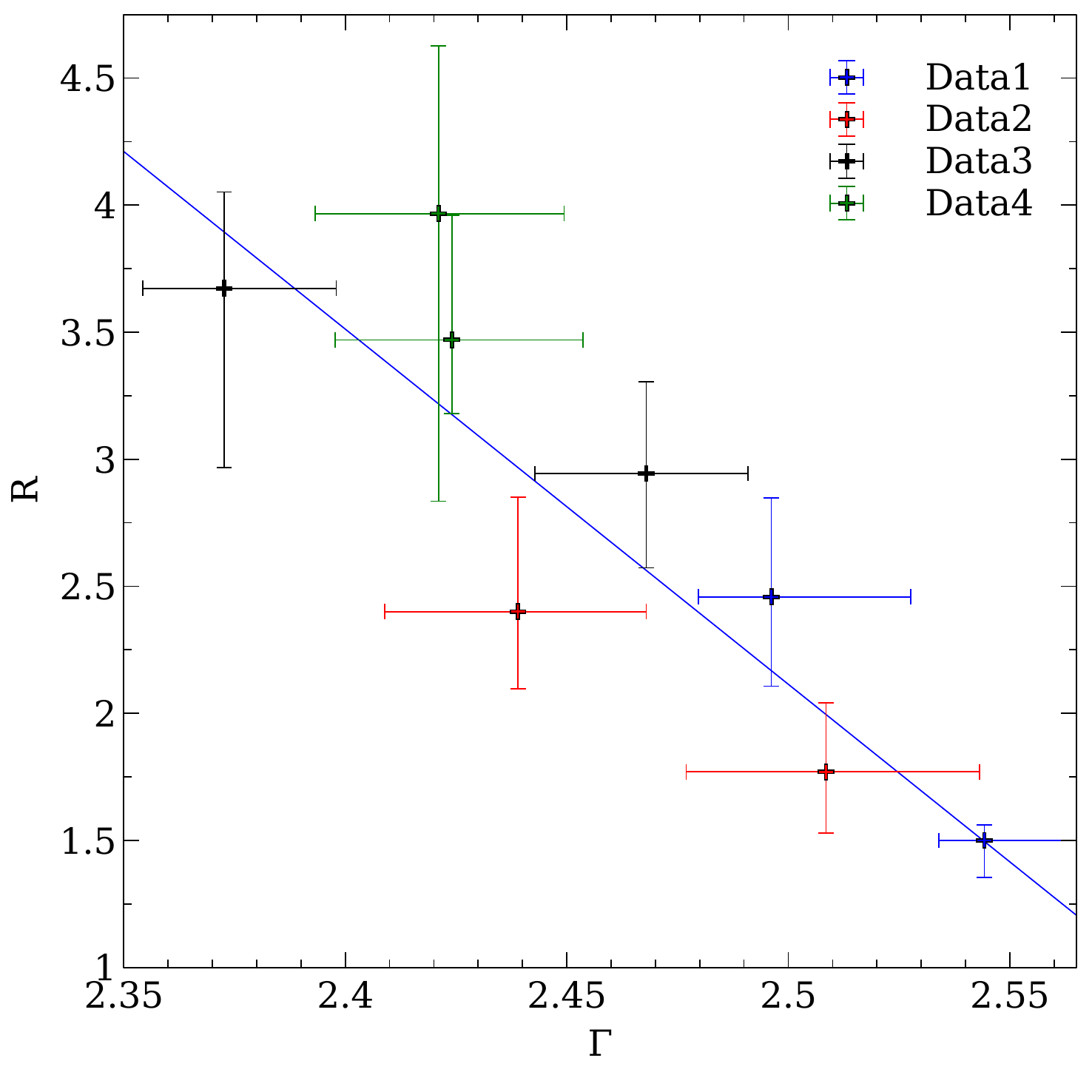}
		\caption{ Plot for reflection fraction (R) versus photon index ($\Gamma$) from the fits of all {\it XMM-Newton} flux-resolved spectra provided by different observations. The blue line is the best fit line obtained from the function $a + bx$, where $a = 37 \pm 5$, $b = -14 \pm 2$ and reduced $\chi^2 = 0.99$.}
		\label{fig:figure4}
	\end{minipage}
\end{figure*}

\subsection{Flux resolved spectra}

Each of the {\it XMM-Newton} observations has significant variability and signal to noise,  to undertake a two level flux-resolved spectroscopy. As a representative case, the lightcurve for Data4 is shown in Figure~\ref{fig:figure2} binned with 500 seconds. Clear variability is seen, allowing us to divide the data into two flux (low and high) states, having roughly the same exposure time, based on the count rate being higher or lower than $\sim 3.6$ counts/s. Spectra for the high and low flux states were extracted and the procedure was repeated for the other four observations.

The same model as used for the time-averaged spectra was used to fit the high and low flux spectra of each of the {\it XMM-Newton} observations. The black hole spin, the inclination angle and the iron abundance were fixed at the values obtained in the time-averaged analysis. The best-fit parameters obtained for each data set have been tabulated in Table~\ref{tab:table5}.

\subsection{Dependence of reflection fraction on flux and index}

From the spectral parameters obtained from the time averaged {\it Swift} data (Table~\ref{tab:table4}) and from the flux-resolved ones listed in Table~\ref{tab:table5}, it can be inferred that the reflection fraction inversely correlates with the flux and index. This is illustrated in  Figure~\ref{fig:figure3} and Figure~\ref{fig:figure4}, where the reflection fraction $R$ is plotted against flux and index, respectively for the flux-resolved analysis. Note that the flux quoted here is the total flux (primary and reflected) from the source, while several earlier works have reported the variation of reflection fraction with the primary or power-law flux. Since the reflection fraction is higher for the lower total flux observations, the primary flux would be proportionally smaller for those observations. Hence the reflection fraction would have an inverse correlation with the  primary flux, similar to the one reported here for the total flux.

\begin{figure}
	\includegraphics[width=8.0cm, height=6.5cm]{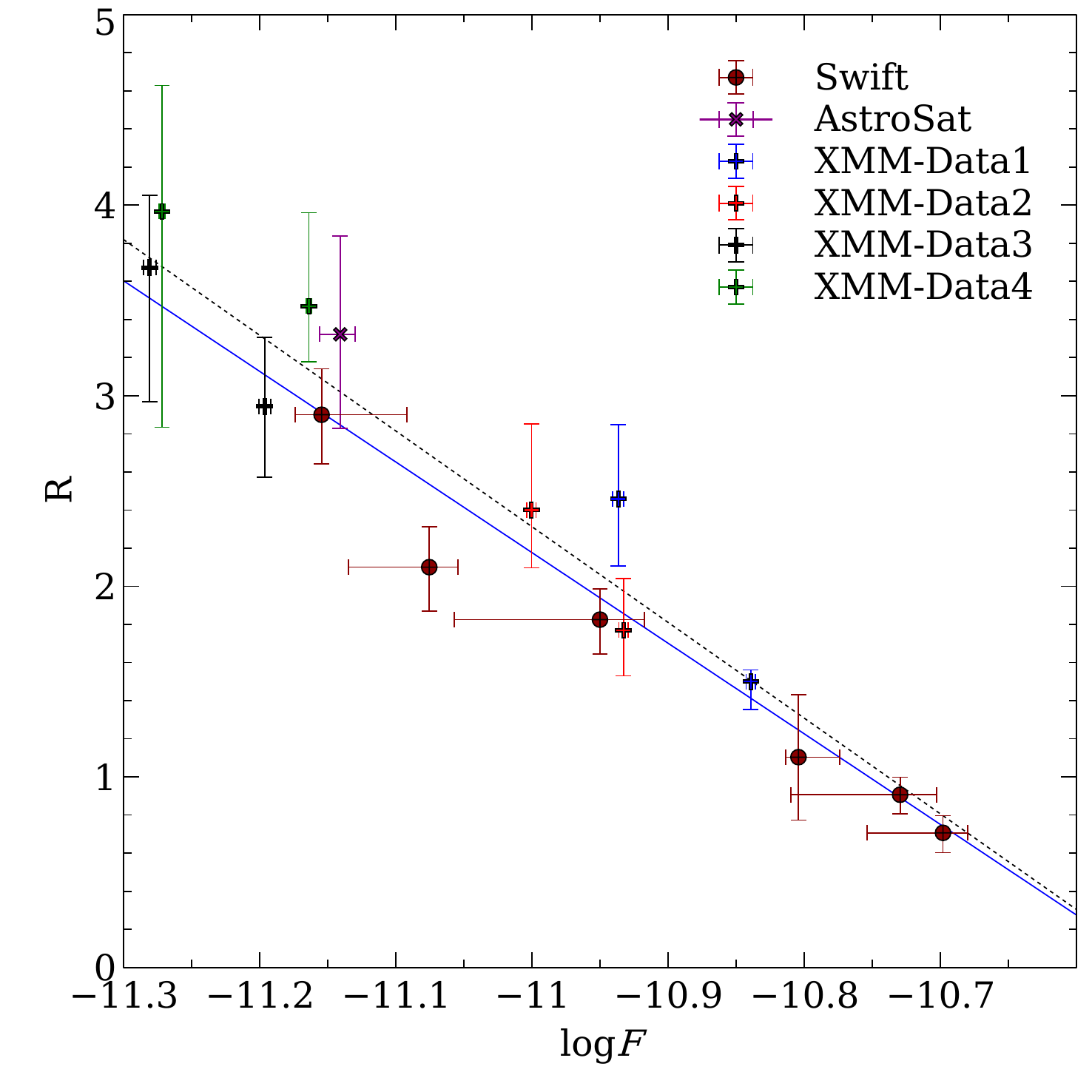}
	\caption{Reflection fraction (R) versus flux obtained from different time-averaged spectra from {\it Swift} and {\it AstroSat}, and flux-resolved {\it XMM-Newton} spectra. The best fit line (blue) is obtained from the function $a + bx$, where $a = -50 \pm 3.32$, $b = -4.75 \pm 0.30$ and reduced $\chi^2 = 0.9$. The dotted line is the best fit line from Figure 3.}
	\label{fig:figure5}
\end{figure}

\begin{figure}
	\includegraphics[width=8.0cm, height=6.5cm]{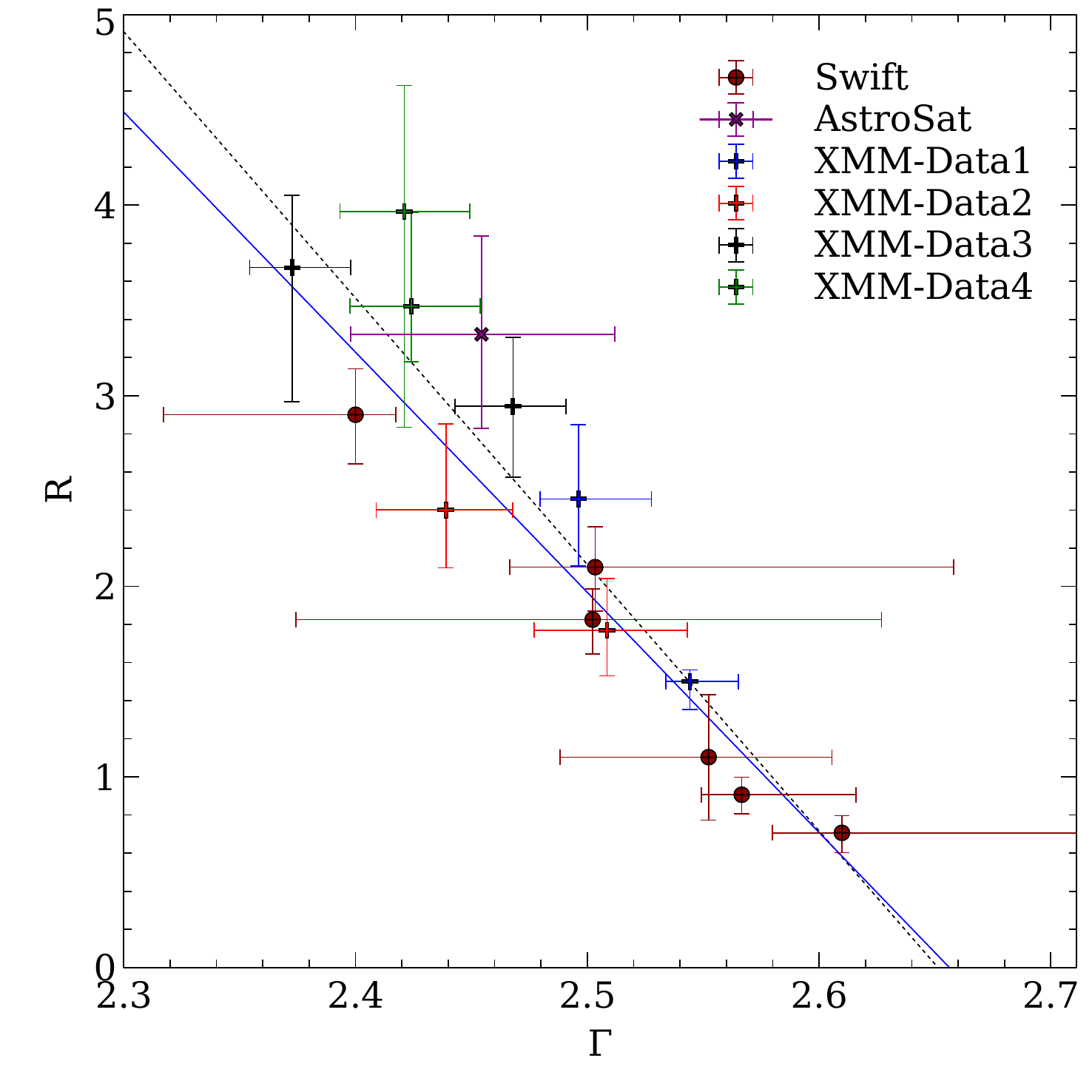}
	\caption{Reflection fraction (R) versus photon index ($\Gamma$) from the fits of all {\it XMM-Newton} flux-resolved spectra and time-averaged {\it AstroSat} and {\it Swift} spectra obtained from different observations. The best fit line (blue) is obtained from the function $a + bx$, where $a= 33.5 \pm 2.66$, $b=-12 \pm 1.04$ and reduced $\chi^2 = 1.5$. The dotted line is the best fit line from Figure 4.}
	\label{fig:figure6}
\end{figure}

It is interesting to note that the general correlation observed for the time-averaged spectra between the $R$, flux and index (i.e. on time-scales of years) holds broadly true when flux-resolved analysis is considered (i.e. on time scales of hours). This suggestion is reinforced when the results from {\it Swift} and {\it AstroSat} analysis is included as shown in 
Figure~\ref{fig:figure5} and Figure~\ref{fig:figure6}. Here the results of one of the {\it Swift} observations, Data9 (obsid: 00091339003), has been omitted since for that data, the reflection fraction was not constrained. In Figure 7, we plot the photon index versus logarithm of the flux which shows a clear positive correlation. The Figures show that the long-term correlation obtained from {\it Swift} and {\it AstroSat} data are similar to the ones obtained from the flux-resolved {\it XMM-Newton} analysis. Thus, the temporal behaviour of the system on years time-scales is similar to the one exhibited on hours time-scales.

\section{Discussions}

We performed spectral analysis of Mkn 478 using {\it XMM-Newton}, {\it AstroSat} and {\it Swift} observations using a blurred relativistic reflection model and as was reported earlier \citep{Zoghbi2008,Waddell2019}, we find that the source is reflection dominated. We report that the reflection fraction inversely correlates with the flux and power-law index over time-scales of years and using flux-resolved spectroscopy for the {\it XMM-Newton} observations, we find that a similar correlation is obtained for time-scales of hours. Moreover, similar variability can also be inferred for intermediate time-scales of days to months which is the time separation for some of the  {\it Swift} and {\it XMM-Newton} observations.

Using six {\it XMM-Newton} observations of 1H 0419-577, \citet{Fabian2005} reported the inverse correlation between the reflection fraction and flux. Here, we have extended the result to Mkn 478 and hence affirm that this inverse correlation occurs for other reflection dominated AGN. The correlation has been explained in the framework of a ``lamp-post'' model
\citep{Miniutti2004, Niedzwiecki2016, Wilkins2021, Zoghbi2021}. In this model, the hard X-ray emitting region is approximated as a point source located above the black hole perpendicular to the accretion disc. When the source is close to the black hole, its intrinsic emission is strongly red-shifted, while light bending effects lead to an enhanced reflection giving rise to the observed reflection dominated spectra. Thus, movement of the X-ray source towards the black hole results in lower flux and higher reflection, thereby explaining the inverse correlation between flux and reflection fraction. Our finding reveals that the same correlation occurs for years and hours time-scales, suggesting that the variability of the source is determined by the movement of the corona over a wide range of time-scales. Such a ``lamp-post'' model has been incorporated as a variation of the the {\tt relxill} family of models, and is called {\tt relxilllp}. Here, the height of the point source X-ray emitting source (above the black hole and perpendicular to the disc) can be fitted as one of the spectral parameters.  Applying this model to the flux resolved spectral analysis of Data3, we find that as expected the height increases from  $2.11_{-0.003}^{+0.015}$ to $2.71_{-0.13}^{+0.21}$ gravitational radii as the source increases in flux. For a  $\sim 2 \times 10^7$ solar mass black hole \citep{Porquet2004}, a $\sim 0.5 r_g$ variation in $\sim 10$ ksecs would imply a speed of $\sim 1.5 \times 10^8$ cm/s or $0.005$ times the speed of light. However, this is based on a point source interpretation and in general the time-scale should be associated with structural changes of an extended corona.

\begin{figure}
	\includegraphics[width=8.0cm, height=6.5cm]{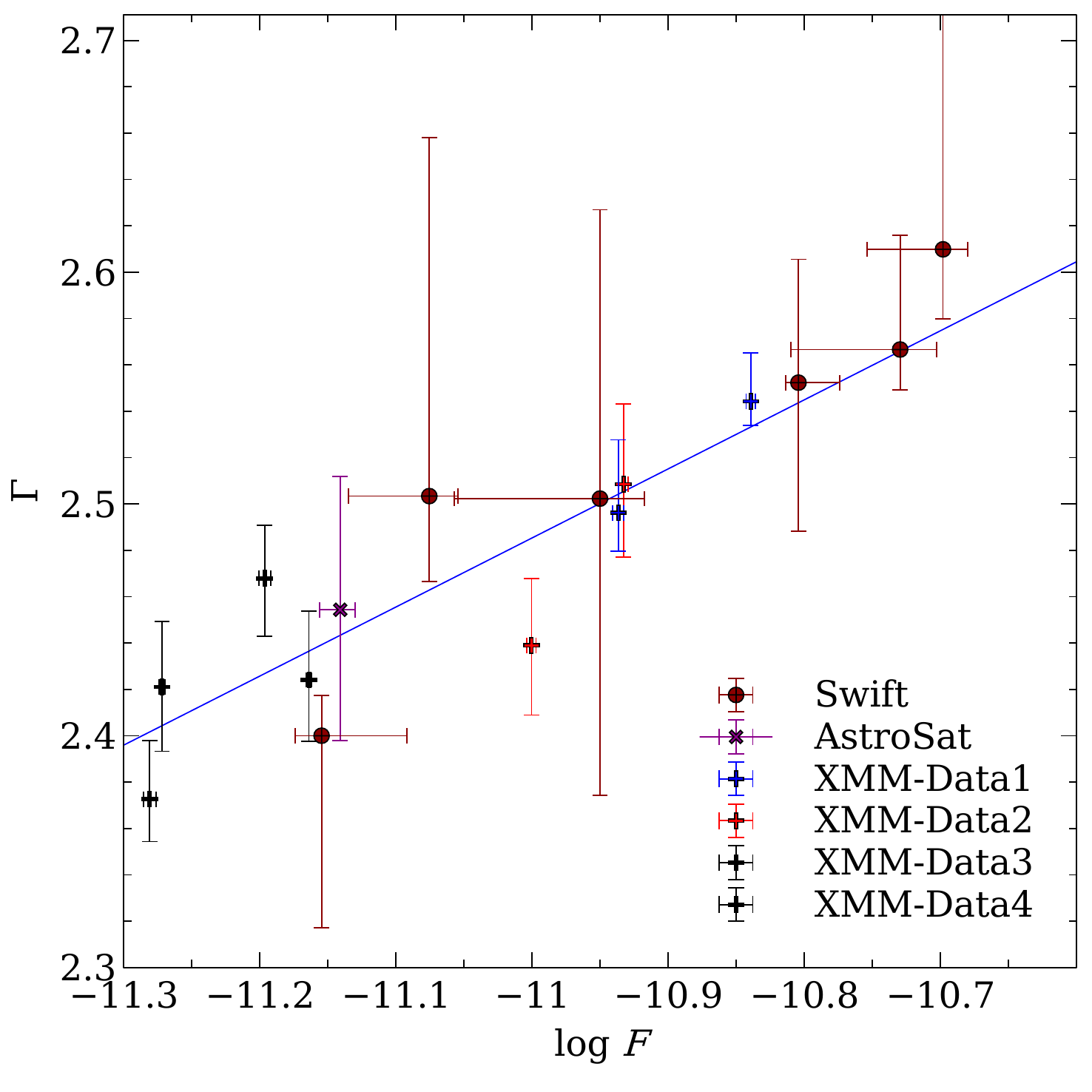}
	\caption{Photon index ($\Gamma$) versus flux from the fits of all {\it XMM-Newton} flux-resolved spectra and time-averaged {\it AstroSat} and {\it Swift} spectra obtained from different observations. The best fit line (blue) is obtained from the function $ a + bx$, where $a = 6 \pm 0.69$, $b = 3 \pm 0.06$ and reduced $\chi^2 = 0.86$.}
	\label{fig:figure7}
\end{figure}

The results presented here show that apart from the flux, the reflection fraction is also inversely correlated with the photon index. 
The inverse correlation between the reflection fraction and the photon index is opposite to what has been reported for regular AGN where the reflection component is not dominant \citep{Dadina2008, Qiao2017, Zappacosta2018, Ezhikode2020}. For those more typical AGN, the positive correlation between reflection fraction and index has been interpreted as a geometry change which leads to a changing solid angle being subtended by the corona for the disc \citep{Zdziarski1999, Boissay2016, Zappacosta2018, Ezhikode2020}. An increased solid angle would lead to a larger reflection fraction and at the same time an increase in the soft photon flux into the corona occurs. If the heating rate of the corona remains same, the increased soft photon flux would lead to a decrease in the coronal temperature which in turn would lead to increase in the photon index, leading to the observed anti-correlation between reflection fraction and photon index. Indeed the decrease in the coronal temperature with increasing index has been detected for AGN \citep{Barua2020} although the opposite effect has also been observed \citep{Barua2021}. As mentioned in the Introduction, for these regular (i.e. not reflection dominated) AGNs, the photon index is correlated to the flux, which is the same behaviour reported here for Mkn 478  (see Figure~\ref{fig:figure7}). Thus, for reflection dominated sources the photon index is correlated with flux as in the case of other AGN, but the variation of the reflection fraction with these two quantities differs.

The results presented here again suggest that for the reflection dominated systems, the scenario is more complex and perhaps as the source moves closer to the black hole (i.e. as the reflection fraction increases), the heating rate of the corona increases, leading to a higher temperature and lower photon index. However, given the spectral quality and complexity of these sources it is not possible to make conclusions and the above inference should be considered speculative. Nevertheless, spectral studies of time varying reflection dominated AGN do provide a good opportunity to unravel the nature of the behaviour of matter close to a black hole.

\section*{Acknowledgements}

We thank the anonymous referee for the constructive comments and suggestions that improved this manuscript. SB acknowledge Inter-University Centre for Astronomy and Astrophysics (IUCAA) for the visiting program and providing technical support to conduct this research. We also thank Prof. Gulab C Dewangan for the useful discussion and support. SB, RM and VJ thank Indian Space Research Organisation (ISRO) for observing the source and providing the data at Indian Space Science Data Centre (ISSDC). This research has also made use of data obtained through the High Energy Astrophysics Science Archive Research Center Online Service, provided by the NASA/Goddard Space Flight Center.

\section*{Data Availability}

The data used in this article are available in the HEASARC database ({\url{https://heasarc.gsfc.nasa.gov}) and in ISSDC ({\url{https://www.issdc.gov.in}).
 



\bibliographystyle{mnras}
\bibliography{reference} 





\bsp	
\label{lastpage}
\end{document}